\documentclass[conference]{IEEEtran}
\usepackage[T1]{fontenc}
\usepackage{cite}
\usepackage{amsmath,amssymb,amsfonts,bm}
\usepackage{algorithm}
\usepackage{algorithmic}
\ifCLASSINFOpdf
	\usepackage[pdftex]{graphicx}
\else
	\usepackage[dvips]{graphicx}
\fi
\usepackage{xcolor}
\usepackage[caption=false,font=footnotesize]{subfig}
\begin{document}
	
	\title{Adaptive Task Partitioning at Local Device or Remote Edge Server for Offloading in MEC}
	
	\author{\IEEEauthorblockN{Jianhui Liu, Qi Zhang}
		\IEEEauthorblockA{DIGIT, Department of Engineering, Aarhus University, Denmark\\
			Email: \{jianhui.liu, qz\}@eng.au.dk}
	}
	
	\maketitle
	
\begin{abstract}
Mobile edge computing (MEC) is one of the promising solutions to process computational-intensive tasks for the emerging time-critical Internet-of-Things (IoT) use cases, e.g., virtual reality (VR), augmented reality (AR), autonomous vehicle. The latency can be reduced further, when a task is partitioned and computed by multiple edge servers' (ESs) collaboration. However, the state-of-the-art work studies the MEC-enabled offloading based on a static framework, which partitions tasks at either the local user equipment (UE) or the primary ES. The dynamic selection between the two offloading schemes has not been well studied yet. In this paper, we investigate a dynamic offloading framework in a multi-user scenario. Each UE can decide who partitions a task according to the network status, e.g., channel quality and allocated computation resource. Based on the framework, we model the latency to complete a task, and formulate an optimization problem to minimize the average latency among UEs. The problem is solved by jointly optimizing task partitioning and the allocation of the communication and computation resources. The numerical results show that, compared with the static offloading schemes, the proposed algorithm achieves the lower latency in all tested scenarios. Moreover, both mathematical derivation and simulation illustrate that the wireless channel quality difference between a UE and different ESs can be used as an important criterion to determine the right scheme.
\end{abstract}
%

\IEEEpeerreviewmaketitle

\section{Introduction}
A variety of the emerging Internet-of-Things (IoT) use cases, e.g., virtual reality (VR), augmented reality (AR), autonomous vehicle, factory automation, and remote surgery etc., require real-time control and steering of cyber physical systems. These use cases often are challenged by processing computation-intensive tasks within the latency constraint of millisecond level \cite{Zhang2015,zhang2018towards}. However, due to limit computation resource, it is difficult for user equipment (UE) to complete a task within the latency constraint merely by local processing.
Mobile edge computing (MEC) is considered as a promising paradigm to fulfill the stringent latency requirement \cite{mao2017survey,elbamby2019wireless}. Compared with the core Cloud server, edge server (ES) can provide computation capability in close proximity to UE.

In the MEC paradigm, computation offloading is one of the key components, which has been studied in many literature.
Work in \cite{lyu2018energy} and \cite{liu2019ar} focused on the optimal strategy to offload the entire task to one ES. Lyu \textit{et al.} \cite{lyu2018energy} jointly optimized the offloading admission decisions and computation resource allocation among users to guarantee delays and energy saving. Liu \textit{et al.} \cite{liu2019ar} maximized reliability of computation offloading subject to latency constraint by dynamic selections of transmission rates and ESs.

Although MEC has potential to enhance the computation capability of UEs, the resource of one ES is limited, compared to that of core Cloud server. However, considering the dense deployment of the future network, it is promising to further shorten the latency by collaboration of multi-ES. In other words, one task can be partitioned into several sub-tasks and computed by multiple ESs in parallel, which is known as task partitioning offloading. In the state-of-the-art work, researchers mainly studied the task partitioning scheme based on a static framework, i.e., partitioning a task at either the local UE or the associated ES.
Work in \cite{Dinh2017,Liu2018,liu2019code} considered to partition a task at UE, and subsequently offload the sub-tasks to different ESs via wireless channel. Dinh \textit{et al.} \cite{Dinh2017} minimized both energy consumption and tasks runtime by coupling sub-task allocation decisions and CPU frequency scaling at UE. To make tradeoff between latency and reliability of computation offloading, Liu \textit{et al.} \cite{Liu2018} proposed heuristic algorithms to allocate and sequentially offload sub-tasks to different ESs. They further improved the task model with the directed acyclic graph and scheduled the sub-task offloading in \cite{liu2019code}.
Other work \cite{chiu2016, chiu2018, wang2016mobile, ren2019collaborative} offloaded the entire task to a primary ES first, at which the task was subsequently partitioned and distributed to other ESs. Regardless of the transmission latency from UEs to the primary ES, Chiu \textit{et al.} proposed algorithms to minimize the latency to cooperatively complete a task by multiple ESs for single-UE \cite{chiu2016} and multi-UE \cite{chiu2018} scenarios, respectively. Ren \textit{et al.} \cite{ren2019collaborative} studied the collaboration between Edge Computing and Cloud Computing, and proposed an optimal task partitioning strategy for latency minimization. Wang \textit{et al.} \cite{wang2016mobile} minimized the latency and energy consumption to offload a task to one ES, and extended the proposed algorithm to multi-ES scenario on the assumption that the primary ES would partition and distribute sub-tasks to other ESs.

The local partitioning schemes in \cite{Dinh2017,Liu2018,liu2019code} offload the sub-tasks separately via the different wireless links between the UE and the ESs, where a poor wireless link between the UE and one of the ESs could affect the overall latency. The other schemes in \cite{chiu2016, chiu2018, wang2016mobile, ren2019collaborative} select the primary ES which has the best wireless link, and offload the entire task to it. Then the primary ES partitions the task and distributes the sub-tasks among multiple ESs via high quality links. However, partitioning at the primary ES does not always outperform the local partitioning schemes, as it takes two hops to transmit the task data to the final computing ESs.
To minimize the overall latency, in this paper, an adaptive offloading framework is proposed in multi-user scenario to select the optimal scheme to process a task in MEC system, i.e., partitioning the task at either the local UE or the primary ES. The key factors that determine the selection, e.g., quality of wireless links and computation resource at ESs, are studied in different scenarios. Based on the proposed framework, the latency to complete a task is minimized by optimizing task partitioning and allocating communication and computation resources among UEs. 

The rest of the paper is organized as follows. Section II describes the system model and problem formulation. In Section III, we investigate the optimal task partitioning and resource allocation strategy. Sections IV presents the numerical results. Finally, Section V concludes the paper.

\section{System Model and Problem Formulation}
In this section, the system model is introduced first, followed by latency analysis on the two static offloading frameworks. The optimization problem is formulated in the end.

\begin{figure}[!t]
	\centering
	\includegraphics[width=0.5\textwidth]{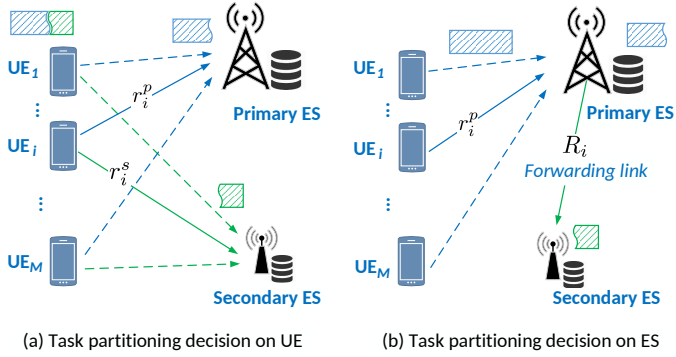}
	\caption{Two static offloading frameworks of task partitioning.}
	\label{fig_off_achi}
\end{figure}

As illustrated in Fig. \ref{fig_off_achi}, we consider an MEC system with collaboration of two ESs. One of the ESs is regarded as the primary ES which is responsible for UE association, control signal exchange and resource scheduling before computation offloading starts, while the other is the secondary ES. Note that the secondary ES is set to share the computation burden of the primary ES and shorten the latency further. The computation resource of the primary and secondary ESs are $f^p$ and $f^s$, respectively. The primary ES is associated with $M$ UEs denoted by a set $\mathcal{M}$, where $\mathcal{M} = \{1,2,\cdots,M\}$. The computation task from UE $i$ is denoted by a tuple $(b_i, \alpha_i)$, where $b_i$ is the input data size of the task (in \textit{bits}) and $\alpha_i$ denotes the required number of CPU cycles for one data bit (in \textit{cycles/bit}). Each task can be computed cooperatively by the two ESs.  Since one computation task in some applications, such as video compression and speech recognition, can be partitioned into independent segments \cite{mao2017survey, ren2019collaborative}, we use $\lambda_i$ to denote the proportion of the sub-task that is assigned from UE $i$ to the primary ES ($ \lambda_i \in [0,1] $). In other word, the primary and secondary ES require to execute sub-tasks of $\lambda_i \alpha_i b_i$ and $(1-\lambda_i) \alpha_i b_i$ CPU cycles for UE $i$, respectively. 

The computation and communication resources are shared by UEs in the MEC system. 
We assume that $f^p_i$ and $f^s_i$ denote the allocated computation resource of the primary and secondary ESs to UE $i$, respectively, where $\sum_{i\in \mathcal{M}}f^p_i \le f^p$ and $\sum_{i\in \mathcal{M}}f^s_i \le f^s$.
UEs access ESs with orthogonal frequency-division multiplexing access (OFDMA) scheme on wireless channel. The total number of resource blocks (RBs) in the MEC system is $N_\textit{rb}$. Given that UE $i$ is assigned with $n_i$ RBs, we have $\sum_{i\in \mathcal{M}}n_i \le N_\textit{rb}$. The available transmission rates (in \textit{bits per second}) of a RB from UE $i$ to the primary and secondary ESs can be adjusted based on channel state information (CSI), which are denoted by $r^p_i$ and $r^s_i$, respectively. In addition, as UEs normally prefer to be associated with a closer ES, we assume $r^p_i \ge r^s_i$.
The task can be also distributed through a forwarding link from the primary ES to the secondary ES as shown in Fig. \ref{fig_off_achi}(b). The forwarding link normally has higher bandwidth than wireless channel. Here we denote the average transmission rate of the link as $R_i$ for UE $i$.
Due to the fact that the size of a task's output is normally much smaller than the input one, the latency caused by downlink transmission is negligible \cite{Dinh2017,Liu2018,liu2019code}.

\begin{figure}[!t]
	\centering
	\includegraphics[width=0.38\textwidth]{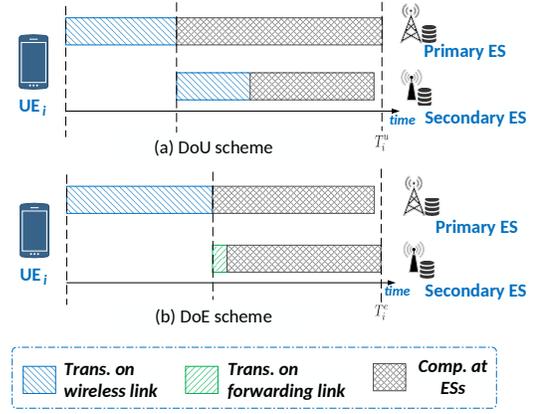}
	\caption{Latency illustration for the two static offloading schemes of  task partitioning.}
	\label{fig_latency_analysis}
\end{figure}

\subsection{Latency Analysis}

To offload computation tasks to the two ESs, each UE has two different options to partition a task: either at the local UE or at the primary ES, as shown in Fig. \ref{fig_off_achi}. The latency illustration of the two schemes is compared in Fig. \ref{fig_latency_analysis} and analyzed as follows.

\subsubsection{Make Task Partitioning Decision on UE (DoU)}
In this scheme, a UE partitions the task into two sub-tasks locally, and subsequently transmits sub-tasks to ESs via wireless channel. In this paper, we consider to transmit sub-tasks sequentially, which is more efficient than offloading in parallel proofed by \cite{Liu2018}. Specifically, UE $i$ offloads the sub-task to the primary ES first, using all the allocated RBs, where the transmission latency is $\frac{\lambda_i b_i}{n_i r^p_i}$. It will start the offloading to the secondary ES after completing the previous transmission. Considering both communication and computation delay, the latency to complete the corresponding sub-task at the primary and secondary ES is $\frac{\lambda_i b_i}{n_i r^p_i} + \frac{\lambda_i \alpha_i b_i}{f^p_i}$ and $\frac{\lambda_i b_i}{n_i r^p_i} + \frac{(1-\lambda_i) b_i}{n_i r^s_i} + \frac{(1-\lambda_i)\alpha_i b_i}{f^s_i}$, respectively. Therefore, the overall latency to complete the task of UE $i$ is
\begin{equation}
	\label{eq_dou_delay}
	T^u_i \! = \! \frac{\lambda_i b_i}{n_i r^p_i} \!+\! \max\left\lbrace \frac{\lambda_i \alpha_ib_i}{f^p_i},  \frac{(1-\lambda_i)b_i}{n_i r^s_i} \! + \! \frac{(1-\lambda_i)\alpha_ib_i}{f^s_i}\right\rbrace.
\end{equation}

\subsubsection{Make Task Partitioning Decision on ES (DoE)}
In this scheme, a UE offloads the entire task to the primary ES first via wireless channel, of which the transmission delay is $\frac{b_i}{n_i r^p_i}$. Subsequently, the primary ES partitions the task and starts to process one part of the task. Meanwhile, the other part of the task is distributed to the secondary ES via the forwarding link. The latency to process sub-task at the primary ES is $\frac{\lambda_i \alpha_i b_i}{f^p_i}$, while the latency introduced by the secondary ES is $\frac{(1-\lambda_i) b_i}{R_i} + \frac{(1-\lambda_i)\alpha_i b_i}{f^s_i}$. Therefore, the overall latency to complete the task of UE $i$ is obtained as
\begin{equation}
	\label{eq_doe_delay}
	T^e_i \!=\! \frac{b_i}{n_i r^p_i} \!+\!  \max\left\lbrace \frac{\lambda_i \alpha_ib_i}{f^p_i}, \frac{(1-\lambda_i)b_i}{R_i} \!+\! \frac{(1-\lambda_i)\alpha_ib_i}{f^s_i} \right\rbrace.
\end{equation}

\subsection{Problem Formulation}
According to \eqref{eq_dou_delay} and \eqref{eq_doe_delay}, the latency to complete a task depends on the available resource, e.g., communication data rate and computation resource, for both DoU and DoE schemes. Each UE should select the scheme properly to minimize the overall latency $T_i$, i.e., $T^*_i = \min\{T^u_i, T^e_i\}$.
Moreover, due to the constrained communication and computation resources in the MEC system, the allocation of RBs and computation resource at ESs has to be optimized among multiple UEs simultaneously. Therefore, we formulate a latency minimization problem as
\begin{subequations}
	\begin{align}
	\mbox{\textbf{P1}:}\ \min\limits_{\bm{x},\bm{\lambda},\bm{n},\bm{f}^p,\bm{f}^s} &\ \ \dfrac{1}{M} \sum_{i\in \mathcal{M}} \beta_i \left[ x_iT^u_i + (1-x_i) T^e_i \right]  \nonumber \\
	\textit{s.t.} &\ \ \sum_{i\in \mathcal{M}}n_i \le N_\textit{rb}, \ n_i>0, \label{Prob-1-a} \\
	&\ \ \sum_{i\in \mathcal{M}}f^p_i \le f^p, \ f^p_i>0, \label{Prob-1-b} \\
	&\ \ \sum_{i\in \mathcal{M}}f^s_i \le f^s, \ f^s_i>0, \label{Prob-1-c} \\
	&\ \ x_i \in \{0,1\}, \ \lambda_i \in [0,1], \ \forall i\in\mathcal{M}, \label{Prob-1-d}
	\end{align}
\end{subequations}
where $\beta_i$ is a positive weight depending on the service priority of the UE $i$ and $\sum_{i\in \mathcal{M}}\beta_i=1$. $\bm{x}$, $\bm{\lambda}$, $\bm{n}$, $\bm{f}^p$ and $\bm{f}^s$ denote the sets of scheme selection indicators, task partitioning ratios, assigned number of RBs and allocated computation resource for UEs in the MEC system, i.e., $\bm{x}=\{x_1, \cdots, x_M\}$, $\bm{\lambda}=\{\lambda_1, \cdots, \lambda_M\}$, $\bm{n}=\{n_1, \cdots, n_M\}$, $\bm{f}^p=\{f^p_1, \cdots, f^p_M\}$ and $\bm{f}^s=\{f^s_1, \cdots, f^s_N\}$, respectively. If UE $i$ chooses the DoU scheme, we have $x_i=1$, otherwise, $x_i=0$. The constraints \eqref{Prob-1-a}-\eqref{Prob-1-c} present the limit of communication and computation resources. Due to the expression of the objective function, the problem \textbf{P1} is a non-convex optimization problem.

\section{Dynamic Task Partitioning Scheme}
In this section, a dynamic task partitioning scheme is proposed. We first analyze the optimal task partitioning strategy, then optimally allocate communication and computation resources to multiple UEs. 
\subsection{Optimal Selection of Task Partitioning}
As the objective function in problem \textbf{P1} is too complicated to be solved directly, we first derive the optimal task allocation for both DoU and DoE schemes, i.e., $\lambda_i^{u*}$ and $\lambda_i^{e*}$, respectively, given the assigned number of RBs, $n_i$, and computation resource, $f^p_i$ and $f^s_i$.

A task is completed cooperatively by the primary and secondary ESs. Intuitively, the latency is minimized when the two ESs can complete the allocated sub-tasks simultaneously \cite{Liu2018}. For instance, according to \eqref{eq_dou_delay}, the optimal task partitioning ratio for DoU scheme can be obtained, when $\frac{\lambda_i \alpha_ib_i}{f^p_i} = \frac{(1-\lambda_i)b_i}{n_i r^s_i} \! + \! \frac{(1-\lambda_i)\alpha_ib_i}{f^s_i}$, i.e.,
\begin{equation}
	\label{eq_opt_lambda_u}
	\lambda_i^{u*} = \frac{1/(n_i r^s_i) + \alpha_i/f^s_i}{1/(n_i r^s_i) + \alpha_i/f^s_i + \alpha_i/f^p_i }.
\end{equation}
Similarly, according to \eqref{eq_doe_delay}, the optimal task partitioning ratio for DoE scheme is
\begin{equation}
	\label{eq_opt_lambda_e}
	\lambda_i^{e*} = \frac{1/R_i + \alpha_i/f^s_i}{1/R_i + \alpha_i/f^s_i + \alpha_i/f^p_i }.
\end{equation}
Substitute \eqref{eq_opt_lambda_u} and \eqref{eq_opt_lambda_e} into \eqref{eq_dou_delay} and \eqref{eq_doe_delay}, respectively, the minimal latency to complete a task for UE $i$ using DoU and DoE schemes is obtained as
\begin{equation}
	\label{eq_opt_delay_u}
	T^{u*}_i \left( \lambda_i^{u*} \right) = b_i \left(\frac{1}{n_i r^p_i} +  \frac{\alpha_i}{f^p_i}\right) \lambda_i^{u*},
\end{equation}
\begin{equation}
	\label{eq_opt_delay_e}
	T^{e*}_i \left( \lambda_i^{e*} \right) = b_i \left(\frac{1}{n_i r^p_i} +  \frac{\alpha_i}{f^p_i} \lambda_i^{e*} \right). 
\end{equation}

To select the optimal scheme between the DoU and DoE schemes, we let $T^{e*}_i - T^{u*}_i = 0$. Therefore, the optimal strategy to partition a task for UE $i$ is obtained as follows.

\textit{Theorem 1 (Task Partitioning Decision on UE or ES)}: Given $n_i$, $f^p_i$ and $f^s_i$, UE $i$ can select task partitioning scheme between the DoU and DoE based on the following conditions:
\begin{itemize}
	\item if $\frac{n_i r^s_i}{R_i} + \eta_i\frac{r^s_i}{r^p_i} - 1 \ge 0$, the DoU scheme is selected, i.e., $x_i=1$;
	\item if $\frac{n_i r^s_i}{R_i} + \eta_i\frac{r^s_i}{r^p_i} - 1 < 0$, the DoE scheme is selected, i.e., $x_i=0$,
\end{itemize}
where $\eta_i = \frac{f^p_i}{\alpha_i R_i} + \frac{f^p_i}{f^s_i} + 1$ and $\eta_i \ge 1$.

\begin{figure}[!t]
	\centering
	\includegraphics[width=0.35\textwidth]{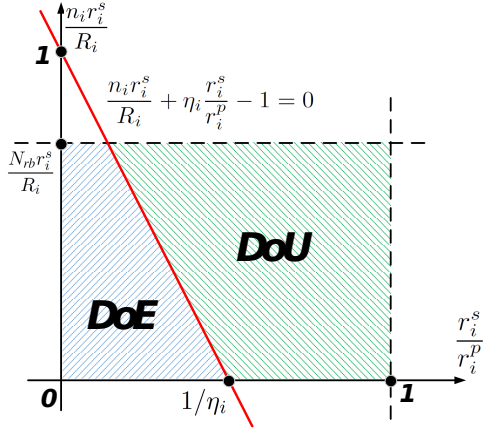}
	\caption{Illustration of optimal selection between DoU and DoE schemes.}
	\label{fig_lemma}
\end{figure}
Note that, as mentioned in the Section II, we have $ 0 < r^s_i / r^p_i\le 1$. In addition, the link capacity between ESs is normally larger than that on wireless channel, i.e., $n_i r^s_i/R_i \le 1$. Therefore, Theorem 1 can be illustrated by Fig. \ref{fig_lemma}. 
It reveals that the optimal task partitioning scheme mainly depends on the $r^s_i/r^p_i$, $n_i r^s_i/R_i$, and $f^p_i/f^s_i$, i.e., the ratio of wireless link quality between the two ESs, the ratio of transmission rate between the wireless channel and the forwarding link, and the ratio of computation resource between the primary and secondary ESs.
If $r^s_i/r^p_i$ is larger than $1/\eta_i$, UE $i$ is preferable to partition a task locally. When $r^s_i/r^p_i < 1/\eta_i$, the selection between the DoE and DoU schemes mainly depends on $n_i r^s_i/R_i$. Moreover, the higher $f^p_i/f^s_i$ leads to the smaller $1/\eta_i$, which means the DoU scheme could outperform DoE scheme with higher probability.

\subsection{Joint Resource Allocation}
So far, we have obtained the optimal task partitioning strategy for UE $i$ when $n_i$, $f^p_i$ and $f^s_i$ are given. In this subsection, we will study the allocation of the communication and computation resource among UEs, i.e., $\bm{n}$, $\bm{f}^p$ and $\bm{f}^s$.

Denoting the optimal latency to complete the task of UE $i$ as $T^*_i$, we have $T^*_i = \min\{T^{u*}_i, T^{e*}_i\}$. Based on the objective function in \textbf{P1} and Theorem 1, the latency of dynamic offloading for UE $i$ could be estimated by
\begin{equation}
	\label{eq_est_opt_t}
	\tilde{T}_i = \frac{S^u_i}{S^u_i + S^e_i}T^{u*}_i + \frac{S^e_i}{S^u_i + S^e_i}T^{e*}_i,
\end{equation}
where $S^u$ and $S^e$ denote the size of trapezoids representing the selection of DoU and DoE schemes in Fig. \ref{fig_lemma}, respectively. We have $\frac{S^e_i}{S^u_i + S^e_i} = \left(2-\frac{n_i r^s_i}{R_i}\right)/(2\eta_i)$ and $ \frac{S^u_i}{S^u_i + S^e_i} = 1 - \frac{S^e_i}{S^u_i + S^e_i}$. Therefore, the \textbf{P1} can be transformed into
\begin{subequations}
	\begin{align}
	\mbox{\textbf{P2}:}\ \min\limits_{\bm{n},\bm{f}^p,\bm{f}^s} &\ \ \frac{1}{M} \sum_{i\in \mathcal{M}} \beta_i \tilde{T}_i \nonumber \\
	\textit{s.t.} &\ \ \eqref{Prob-1-a}-\eqref{Prob-1-d}.
	\end{align}
\end{subequations}
Relaxing $\bm{n}$ from integer space to real number space, the optimal $\bm{n}$, $\bm{f}^p$ and $\bm{f}^s$ in \textbf{P2} can be searched by the interior-point algorithm (IPA). Based on the allocation results, the optimal task partitioning scheme can be selected according to Theorem 1.  

\subsection{Compatibility with Generic Scenarios}
The proposed dynamic task partitioning scheme can be further extended to more general scenarios. We assume the number of secondary ESs is $N$. When $N>1$, the optimal task partitioning ratios for the DoU and DoE schemes are
\begin{align}
	&\lambda_i^{u*} = \frac{\lambda_{i,s}^u \left[ 1/(n_i r^s_{i,1}) + \alpha_i/f^s_{i,1}\right] }{\lambda_{i,s}^u \left[1/(n_i r^s_{i,1}) \! + \! \alpha_i/f^s_{i,1}\right] \! + \! \alpha_i/f^p_i }, \label{eq_opt_lambda_u_ge} \\
	&\lambda_i^{e*} = \frac{\lambda_{i,s}^e \left( 1/R_i + \alpha_i/f^s_{i,1}\right) }{\lambda_{i,s}^e \left(1/R_i + \alpha_i/f^s_{i,1}\right) + \alpha_i/f^p_i }, \label{eq_opt_lambda_e_ge}
\end{align}
where $r^s_{i,j}$ and $f^s_{i,j}$ are the assigned RBs and computation resource to the secondary ES $j$ $(j\le N)$, respectively. $\lambda_{i,s}^u$ and $\lambda_{i,s}^e$ are presented as
$
\left( 1 \!+\! \sum_{j=2}^{N} \frac{\prod_{k=1}^{j-1}\alpha_i/f^s_{i,k}}{\prod_{k=2}^{j}\left[1/(n_i r^s_{i,k}) + \alpha_i/f^s_{i,k}\right]} \right)^{\!-1}
$, 
$
\left( 1 \!+\! \sum_{j=2}^{N} \frac{\prod_{k=1}^{j-1}\alpha_i/f^s_{i,k}}{\prod_{k=2}^{j}\left(1/R_i + \alpha_i/f^s_{i,k}\right)} \right)^{\!-1}
$, respectively.
Based on \eqref{eq_opt_lambda_u_ge} and \eqref{eq_opt_lambda_e_ge}, $f^s_i$ and $r^s_i$ in \eqref{eq_opt_lambda_u} and \eqref{eq_opt_lambda_e} can be replaced by
\begin{align}
	&\tilde{f}^s_i = \alpha_i \left[ \lambda_{i,s}^e \left(1/R_i + \alpha_i/f^s_{i,1}\right) - 1/R_i\right]^{-1}, \label{eq_es_f}\\
	&\tilde{r}^s_i = \left[ \lambda_{i,s}^u \left(1/r^s_{i,1} + n_i\alpha_i/f^s_{i,1}\right) - n_i\alpha_i/\tilde{f}^s_i\right]^{-1}. \label{eq_es_r}
\end{align}
In this way, the optimal scheme selection and resource allocation can be obtained following Section III-A and III-B.

\begin{table}[!t]
	\renewcommand{\arraystretch}{1.3}
	\caption{ Default Simulation Parameters }
	\label{table_sim para}
	\centering
	\begin{tabular}{c c}
		\hline
		Parameter & Value\\
		\hline
		Wireless channel bandwidth & 20 MHz \\
		The number of resource blocks $N_\textit{rb}$ & 100 \\
		Received SNR at ESs & $ [0,30] $ dB \\
		Computing capacity at ESs $f^p$, $f^s$ & $ [2,3]\times10^{10} $ cycles/s \\
		Forwarding link capacity between ESs $R_i$ & $ [50,100] $ Mbps \\
		Task data size $b_i$ & $ [0.8,1.2] $ Mbits \\
		Computation intensity $ \alpha_i $ & 248 cycles/bit \\		
		\hline
	\end{tabular}
\end{table}

\section{Numerical Results}

In this section, we present the simulation results to study the performance of the proposed algorithms.
The bandwidth of wireless channel is 20 MHz with 100 RBs in the MEC system. The signal-to-noise ratio (SNR) between UEs and ESs is obtained by a uniform distribution, ranging from 0 dB to 30 dB. Based on different SNRs, UEs dynamically adjust the modulation and coding scheme (MCS) to guarantee a constant BLER of $10^{-3}$. The SNR-MCS mapping is achieved by the link abstraction model proposed by \cite{mezzavilla2012}. Various MCSs are considered combining different modulations (QPSK, 16QAM, 64QAM) and varying coding rates ranging from 1/9 to 9/10.
The computation resource of ESs ranges from $ 2 \times 10^{10} $ cycles/s to $ 3 \times 10^{10} $ cycles/s \cite{ren2019collaborative}. The forwarding link capacity between the two ESs follows a uniform distribution with $R_i\in[50,100]$ Mbps.
The task data size of UEs ranges from 0.8 Mbits to 1.2 Mbits (e.g. a VR-video frame of 4K resolution is $4096\times2160$ pixels with 24 bit/pixel, using a compression ratio of 300:1 \cite{bastug2017}). The computation intensity of tasks $\alpha$ is set to 238 cycles/bit \cite{miettinen2010}.
The default simulation parameters are summarized in Table \ref{table_sim para}. The numerical results in this section are based on an average value over 5000 Monte Carlo simulations.

We present the average gains of latency from the proposed algorithm to the static DoU and DoE schemes, i.e., $1 - \frac{T^\textit{Prop.}}{T^\textit{DoU}}$ and $1 - \frac{T^\textit{Prop.}}{T^\textit{DoE}}$, respectively. In addition, the DoU and DoE schemes mean all the UEs in the system partition task locally and remotely, respectively. To illustrate the necessity of the proposed scheme, four different scenarios are studied as follows.

\subsection{High Forwarding Link Capacity}
\begin{figure}[!t]
	\centering
	\includegraphics[width=0.38\textwidth]{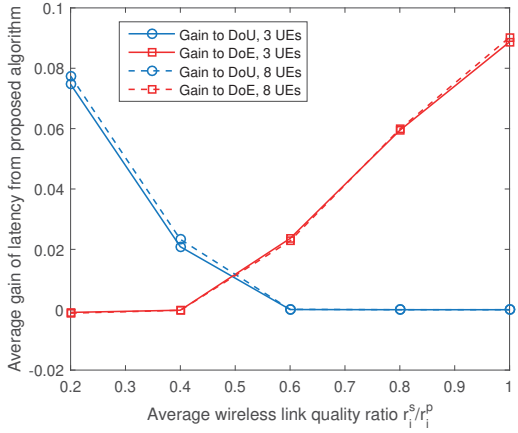}
	\caption{Impact of the number of UEs in high forwarding link capacity scenario.}
	\label{fig_backhaul_Xr_Ylatency_varUE}
\end{figure}

In this scenario, we consider the capacity of the forwarding link is much higher than that of the wireless link, i.e., $R_i \gg N_{rb}r^p_i$. The data rate on the forwarding link $R_i$ is set to 1 Gbps.
Fig. \ref{fig_backhaul_Xr_Ylatency_varUE} shows the latency comparison of different offloading schemes with varying average ratio of wireless link quality between the two ESs, $r^s_i/r^p_i$. The curves in the figure illustrates the average gain of latency from the proposed algorithm to the other two schemes. Since the average latency gain is higher than zero in different settings, the proposed algorithm outperforms both DoU and DoE schemes. Compared with the DoU scheme, the DoE scheme achieves lower latency if the average $r^s_i/r^p_i$ is smaller than 0.5. However, with $r^s_i/r^p_i$ increasing, the latency of two schemes goes in opposite directions. We call the point that two schemes obtain the same latency as the \textit{inflection point}, which approximates to  $r^s_i/r^p_i=0.5$ in this figure. The inflection point refers to a selection indicator of the proposed dynamic offloading algorithm. The inflection point with smaller $r^s_i/r^p_i$ implies DoU scheme has a higher probability to outperform DoE scheme. 
Moreover, the number of UEs in the MEC system have little effect on the average latency.

\begin{figure}[!t]
	\centering
	\includegraphics[width=0.38\textwidth]{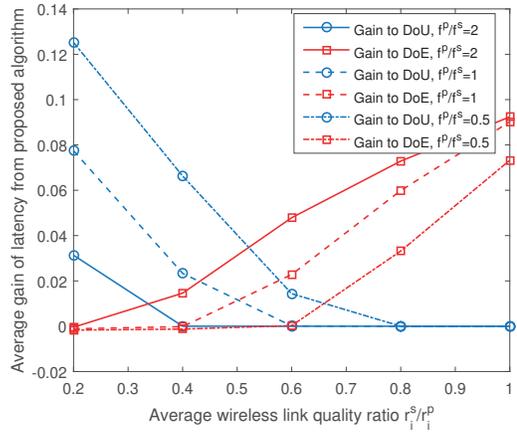}
	\caption{Impact of computation resource ratio between ESs on latency.}
	\label{fig_backhaul_Xr_Ylatency_varCmp}
\end{figure}

In Fig. \ref{fig_backhaul_Xr_Ylatency_varCmp}, we investigate the gains of the proposed algorithm with varying average $f^p/f^s$. The number of UEs here is set to 8. If $r^s_i/r^p_i$ is larger than 0.7, it is preferred to partition task at local UE, regardless of $f^p/f^s$. Moreover, with the average $f^p/f^s$ decreasing, the average $r^s_i/r^p_i$ of the inflection points increases. The reason lies in that more proportion of a task tends to be offloaded to the secondary ES, if the computation resource of the primary ES reduces.

\subsection{Limited Computing Capacity at the Primary ES}
\begin{figure}[!t]
	\centering
	\includegraphics[width=0.38\textwidth]{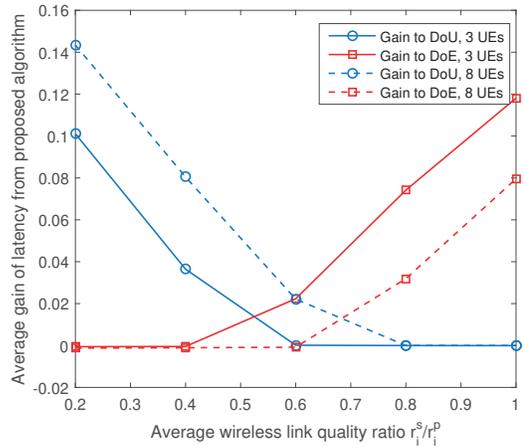}
	\caption{Latency comparisons if computation resource of primary ES is limited.}
	\label{fig_master_Xr_Ylatency}
\end{figure}

This scenario illustrates the impact on the latency to complete a task if the computation resource at the primary ES is much lower than that at the secondary ES. Here $f^p/f^s$ is set to 0.2. Fig. \ref{fig_master_Xr_Ylatency} shows that the proposed algorithm obtains the optimal latency. The gain of the proposed algorithm to the DoU scheme increases with the number of UEs increasing, while the gain to the DoE scheme shows the opposite tendency. Moreover, if the number of UEs grows, more task data will be offloaded to the secondary ES, of which channel quality has more effect on the optimal scheme selection. As a result, the inflection point with 8 UEs is at the right side of that with 3 UEs.


\subsection{Communication Latency Dominated}
\begin{figure}[!t]
	\centering
	\includegraphics[width=0.38\textwidth]{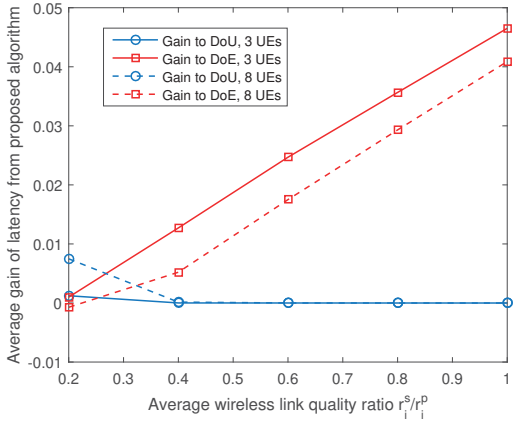}
	\caption{Latency comparisons if communication latency is dominated.}
	\label{fig_comm_Xr_Ylatency}
\end{figure}

The communication latency dominated scenario means the latency to transmit a task on wireless channel is much larger than that to compute the task at ESs. It happens when the communication resource is limited. In this scenario, we assume the computation intensity of a task is 45 cycles/bit (e.g., application of gzip compression \cite{miettinen2010}), which leads to the average ratio of the wireless transmission time to the computation time is around 6:1. In Fig. \ref{fig_comm_Xr_Ylatency}, it shows the average gain of latency from the proposed algorithm to the other two schemes in different network settings. We observe that, if the number of UE is 8, the average $r^s_i/r^p_i$ of the inflection points is close to 0.3. When there are 3 UEs in the MEC system, the DoU scheme will be the better choice regardless of the wireless channel quality ratio. It is because the DoE scheme has to transmit the whole task to ES first via wireless channel. However, it is also worth noticing that, compared with the results in Fig \ref{fig_backhaul_Xr_Ylatency_varUE} and Fig. \ref{fig_master_Xr_Ylatency}, the maximal gains of the proposed algorithm here is smaller, which is less than 5\% for the case $r^s_i/r^p_i=1$. It means that DoU and DoE have approximate latency performance in this scenario.


\subsection{Computation Latency Dominated}
\begin{figure}[!t]
	\centering
	\includegraphics[width=0.38\textwidth]{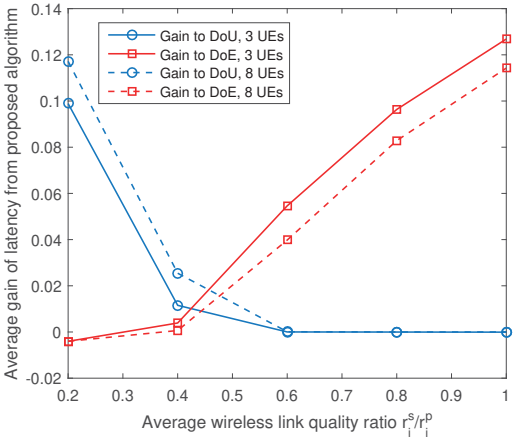}
	\caption{Latency comparisons if computation latency is dominated.}
	\label{fig_comp_Xr_Ylatency}
\end{figure}

In this scenario, the latency to compute the task dominates in the overall latency. We consider the computation resource of each ES follows a uniform distribution ranging from $ 6 \times 10^9 $ cycles/s to $ 10 \times 10^9 $ cycles/s, where the average latency ratio of the communication to the computation is about 1:3. In Fig. \ref{fig_comp_Xr_Ylatency} the inflection points of the cases with different UEs are close to each other. The DoU schemes will achieve lower latency if $r^s_i/r^p_i > 0.5$, whereas it is preferable to partition task at the primary ES if $r^s_i/r^p_i < 0.45$.

\section{Conclusion}
In this paper, we study the optimal offloading scheme of task partitioning for time-critical services in the MEC system. We propose an algorithm to dynamically select the schemes between DoU and DoE for latency minimization. It optimizes the task partitioning, as well as the allocation of communication and computation resource in multi-UE scenario. The numerical results show that, compared with the static DoU and DoE schemes, the proposed algorithm achieves lower latency in various scenarios. It is applicable to the design of the dynamic offloading framework in MEC system. The ratio of wireless channel quality between ESs, i.e., $r^s_i/r^p_i$, can be used as a key criterion to determine the selection of task partitioning. It is preferred to partition a task at the primary ES if $r^s_i/r^p_i \ll 1$. With $r^s_i/r^p_i$ increasing, the latency of DoE scheme gradually exceeds that of DoU scheme. The other factors, e.g., the ratio of computation resource between ESs, as well as the number of UEs, mainly affect the inflection point that indicates the conversion of the better selection at certain $r^s_i/r^p_i$.

\bibliographystyle{IEEEtran}
\bibliography{IEEEabrv,Ref_p5}

\end{document}